\newcommand{\CLASSINPUTbottomtextmargin}{2.5cm}
\documentclass[conference]{IEEEtran}

\usepackage{booktabs}   
\usepackage{subcaption} 

\pdfmapfile{+txfonts.map}
\usepackage{graphicx}
                     
\usepackage{tabularx}
\usepackage{amssymb}
\usepackage{amsmath}
\usepackage{amsthm}
\usepackage{fancyvrb} 

\usepackage[font={small}]{caption}
\usepackage{flushend}

\usepackage{enumitem} 
\usepackage[T1]{fontenc}
\usepackage{listings}
\usepackage[]{algorithm2e}
\usepackage{cite}

\theoremstyle{definition}

\newcolumntype{L}{>{\centering\arraybackslash}m{1.5cm}}

\begin{document}

\title{Optimal and Heuristic Min-Reg Scheduling Algorithms for GPU Programs}         


\author{
\IEEEauthorblockN{Gang Chen}
\IEEEauthorblockA{Intel Corporation, Santa Clara, CA, USA
   \\\ gang.y.chen@intel.com}
}



\maketitle

\setlength{\belowcaptionskip}{-4pt}
\setlength{\abovecaptionskip}{0	pt}
\setlength{\textfloatsep}{6pt plus 1.0pt minus 2.0pt}

\begin{abstract}
Given a basic block of instructions, finding a schedule that requires the minimum number of registers for evaluation
is a well-known problem. The problem is NP-complete when the dependences among instructions form 
a directed-acyclic graph instead of a tree. We are striving to find efficient approximation algorithms 
for this problem not simply because it is an interesting graph optimization problem in theory. 
A good solution to this problem is also an essential component in solving the more complex instruction scheduling problem on GPU.

In this paper, we start with explanations on why this problem is important in GPU instruction scheduling. 
We then explore two different approaches to tackling this problem.
First we model this problem as a constraint-programming problem. 
Using a state-of-the-art CP-SAT solver, we can find optimal answers for much larger cases than previous works on a modest desktop PC. 
Second, guided by the optimal answers, we design and evaluate heuristics
that can be applied to the polynomial-time list scheduling algorithms. A combination of those heuristics 
can achieve the register-pressure results that are about 17\% higher than the optimal minimum on average. However, there are still near 6\% cases 
in which the register pressure by the heuristic approach is 50\% higher than the optimal minimum.
\end{abstract}

\section{Introduction}
\label{sec:intro}
Given a basic block of instructions, we want to find one sequential order that
satisfies all the data-dependence constraints yet minimizes the number of registers required.
In this paper, we call this problem the min-reg scheduling problem.
A good algorithm to min-reg scheduling has multiple practical applications. For example, many modern GPUs
come with compute-core design in which the register file can be flexibly allocated among multiple active threads. 
The number of active threads running on one core is limited by the number of registers each thread requires. 
When compiler generates code that uses less registers, more active threads can be launched in parallel to 
achieve higher computing throughput. Of course, simply minimizing registers is not always the best performing solution. 
The number of active threads can be limited by other hardware resources. 
In reality, we need to pick the best performing solution among a spectrum of solutions: 
between a solution that maximizes thread-level parallelism to a solution that maximizes instruction-level parallelism. 
The min-reg schedule is important because it sits at one end of this spectrum. 
In GPU instruction scheduling, we can use it as the starting point of searching the spectrum.
Even on GPUs with a fixed number of registers per thread, one practical approach to implement a register-limited prepass scheduler 
is using two phases: first a min-reg scheduler then a register-limited latency-hiding scheduler. 
The register pressure is kept under the limit for a greedy latency-hiding scheduler by subdividing a basic block into sub-blocks.
The outcome of this subdividing algorithm is based upon the instruction order after min-reg scheduling. 
A better min-reg schedule helps create larger sub-blocks for latency-hiding scheduling. 
Readers who are interested can refer to~\cite{IGC-CGO} for details.

Also modern CPUs use hardware instruction scheduler to exploit 
instruction-level parallelism. Therefore there is less emphasis on compile-time scheduling for latency hiding. 
Instead the compiler needs to focus on how to utilize the architectural registers efficiently, and minimize spill code.  

\begin{figure}[h]
\center
\includegraphics[width=0.3\textwidth]{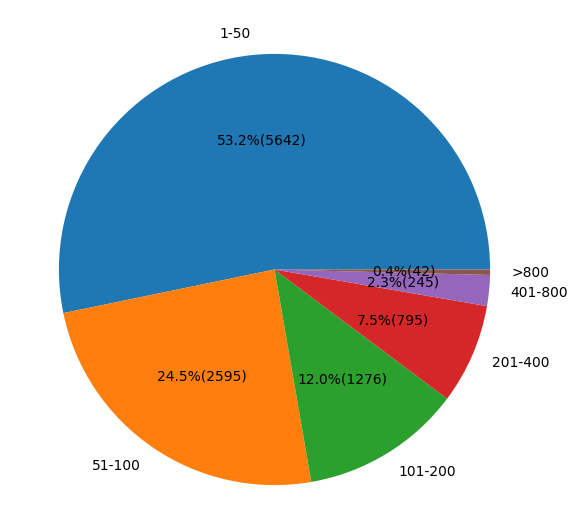}
\caption{Distribution of Basic Block Size in Pixel Shaders}
\label{fig:pie}
\end{figure}

However an efficient algorithm to find the optimal min-reg schedule only exists 
when the data-dependence graph formed by instructions is a tree. In other words, 
the output of every instruction can only have single use. See~\cite{Sethi:1970:GOC:321607.321620} and~\cite{appel1987generalizations}.
Looking at the min-reg scheduling problem from a different angle, we can model it 
as an integer or constraint programming problem. Even though modeling the problem 
as an operations research (OR) problem does not change its NP-completeness, the 
recent advance of OR tools enables us to find the optimal solution for much larger cases
with a very modest amount of computing power. 
We are looking for effective min-reg scheduling algorithms for basic blocks with hundreds of instructions 
because basic blocks of such size are very common in GPU programs as the results of aggressive function inlining and loop unrolling.
 Figure~\ref{fig:pie} shows the distribution of basic-block sizes in a collection of pixel-shaders.  
Larger basic blocks tend to need a higher number of registers. The need to find the min-reg schedule for them is therefore more acute. 

In this paper, we first present a constraint-programming model for the min-reg scheduling problem. 
Using Google OR-Tools~\cite{ortools}, we can find optimal solutions
for basic blocks with several hundreds instructions on a 4-core Intel I7-7700 desktop within two hours.
With the guidance of the optimal solutions, we explore heuristics that we can apply to
a list scheduler for finding the min-reg schedule. All the heuristics that we have considered are practical to implement in production compilers.
In this paper, we incorporate these heuristics into the list scheduler one-by-one, and present the experimental results after adding each heuristic.

The rest of the paper is organized as follows. 
Section~\ref{sec:overview} gives a brief overview of the min-reg scheduling problem and the major previous works on it.
Section~\ref{sec:cp-sat-solver} describes in detail how to model the min-reg scheduling problem as a constraint-programming problem.  
Section~\ref{sec:scheduler} presents the heuristics that are effective in building  a min-reg list scheduler. 
Section~\ref{sec:exp} shows experimental results of applying our algorithm in a graphics compiler. 
Section~\ref{sec:related} presents related works and Section~\ref{sec:conclusion} concludes the paper.

\section{Problem Overview}
\label{sec:overview}

The execution constraints among a list of instructions can be described using a directed-acyclic graph (DAG) with nodes representing instructions and edges representing data dependencies.
Every node consumes some source operands stored in register file, then produces some destination operands that are also stored in register file.
A linear order of DAG nodes that satisfies all the constraints is called an instruction schedule.
For a given schedule of N nodes, there are N time-steps, one node per time-step. 
At any time-step, we need a certain amount of storage in the register file to store all the operands that are still needed for later usage.  
We say an operand is \textit{live} at time-step \textit{t} when its producer is scheduled before time-step \textit{t}, and at least one of it consumer is scheduled at or after time-step \textit{t}.
The total amount of register-file storage needed at time-step \textit{t} is called the register pressure at time-step \textit{t}, $RP_t$.
The maximum register pressure of a given schedule, \textit{MaxRP}, is the maximum of all $RP_t$ for all \textit{t} from 1 to N.
The goal of the min-reg scheduling is to find the schedule with the minimum \textit{MaxRP}.

There is an efficient algorithm for the min-reg scheduling problem when the DAG is limited to be a computation tree.
The initial algorithm for arranging computation trees was proposed by Sethi and Ullman. 
That algorithm has several limitations~\cite{Aho:1977:CGE:321992.322001}:
The tree must be a binary tree; Each internal node has exactly two child nodes;
Each node must output exactly one register; Finally all registers must be the same size.
Later one, Appel and Supowit extended the algorithm to handle the more general computation trees~\cite{appel1987generalizations}, 
allowing instructions with arbitrary number of source operands and allowing instructions with arbitrary destination size. 

To summarize the generalized Sethi-Ullman algorithm by Appel and Supowit in a very brief way:
\begin{itemize}
\item {It works for trees because a tree can be subdivided into disjoint subtrees.}
\item {The optimal solution for a tree is derived from the optimal solutions of its subtrees.}
\end{itemize}
Essentially the generalized Sethi-Ullmman algorithm is a dynamic-programming algorithm. Let's denote the $MaxRP_i$ as the
the maximum registers required to evaluate all children of tree node \textit{i}, $DefSZ_i$ as the output size of tree node \textit{i}.
Deriving $MaxRP_i$ from its children's results is a two-step approach:
\begin{itemize}
\item {sort the children of node \textit{i} into a descending order so that \[MaxRP_{c_1} - DefSZ_{c_1} \geq\]\[ MaxRP_{c_2} - DefSZ_{c_2}\geq\]\[ ... \geq MaxRP_{c_k} - DefSZ_{c_k}\]}
\item {compute $MaxRP_i$ as follows \[MaxRP_i = max(MaxRP_{c_1}, DefSZ_{c_1}+ \]\[max(MaxRP_{c_2}, DefSZ_{c_2}+max(...,\]\[ DefSZ_{c_{(k-1)}}+max(DefSZ_{c_k}, MaxRP_{c_k}))...)\]}
\end{itemize}
The sorting order implies all instructions in SubTree($c_1$) are executed before all instructions in SubTree($c_2$) and so on.

The generalized Sethi-Ullman algorithm cannot produce the optimal solution on an instruction sequence that forms a more general directed-acyclic graph.
The fundamental difference between a DAG and a tree is that the output of a DAG node can be used by multiple nodes. 
Therefore, a DAG cannot be subdivided into disjoint subtrees on a single def-use edge per subtree.

The min-reg scheduling for DAG is NP-complete. Govindarajan \textit{et al}~\cite{Govindarajan:2003:MRI:642768.642771} proposed a way to model the problem as an integer-programming problem.
Their integer-programming model starts with defining integer variables $f_1$, ..., $f_n$ representing the time-step of each instruction. 
Modeling the data-dependence constraints using $f_1$, ..., $f_n$ is relatively straightforward. 
The challenging parts are on modeling the live-ranges and deriving the register pressure at each time-step.
We would not go into those details of their model. Using their integer-programming model, they reported solving the min-reg scheduling problem with very limited size 
(median of 10 nodes, a geometric mean of 12 nodes, and an arithmetic mean of 19 nodes). 
The size limit may be related to their model design. 
It could also be partially due to the computation power available at that time and the integer-programming tools available at that time.

We propose a different model to formulate the min-reg scheduling problem as a constraint-programming problem. 
Starting with a matrix of boolean variables to describe the schedule, we can come up with a much simpler model with all boolean variables except the objective function.
With the aid of award-winning Google OR-Tools, our model can find optimal solutions for the min-reg scheduling problems of much larger sizes.
Our data set includes 4437 DAGs that can be solved optimally using our CP-SAT solver, in which the smallest DAG has 48 nodes, the largest DAG has 511 nodes, and the average is around 90 nodes. Another significant difference is that our DAGs are all from GPU programs while DAGs in the previous work are all from CPU programs. In the next section, we describe this CP-SAT model in detail.

\section{CP-SAT Solver}
\label{sec:cp-sat-solver}

The CP-SAT formulation of the min-reg scheduling problem involves several $N\times N$ matrices of boolean variables.
First we define a matrix of boolean $x_{ik}$ that describes the schedule.
When $x_{ik}$ is set to one, it means node \textit{i} is scheduled at time-step \textit{k}.
The constraints are there should be only a single bit set to one per row and per column.
\[\sum_{i=1}^{n} x_{ik} \equiv 1\]
\[\sum_{k=1}^{n} x_{ik} \equiv 1\] 
To model all the ordering constraints, we need to introduce a sequence of intermediate integer variables $t_i$,
which means node \textit{i} is scheduled at time-step $t_i$. $t_i$ can be derived from a dot-product of row $x_i$ with the constant vector of (1, ..., n)
\[t_i = \sum_{k=1}^{n} x_{ik}\times k\]
Hence an ordering constraint dictates node \textit{i} must be executed before \textit{j} can be added as
\[t_i < t_j\]
To avoid intermediate integer variables, we do not really introduce $t_i$ in the implementation. We simply add the inequality between two dot-products.
\[\sum_{k=1}^{n} (x_{ik} - x_{jk})\times k < 0\]

\begin{figure}[h]
\includegraphics[width=\columnwidth]{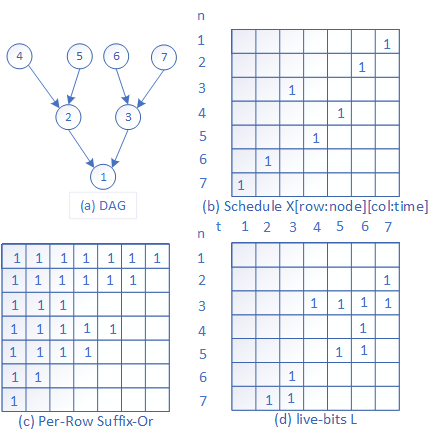}
\caption{Three matrices of boolean variables used in CP-SAT solver}
\label{fig:cp}
\end{figure}

Next, we define a matrix of boolean $ox_{ik}$, in which every element is a suffix-or of $x_{ik}$ per row.
\[ox_{ik} = x_{ik}\text{ when } k \equiv N\]
\[ox_{ik} = x_{ik} \lor ox_{i(k+1)} \text{ when } k < N\]
Defining $ox_{ik}$ is the critical step towards modeling the live-range of node \textit{i}
When the output of node \textit{i} is live at time-step \textit{k} because it is used by node \textit{j},
the following equality must hold:
\[ox_{jk} - ox_{ik} \equiv 1\]
With that, we can derive a matrix of live-bit $l_{ik}$, which stands for the output of node \textit{i} is live at time-step \textit{k}
\[l_{ik} = \lor_{j \in uses(i)} (ox_{jk} - ox_{ik})\]
 In Figure~\ref{fig:cp}, we use a simple example to illustrate the aforementioned process of deriving the live-bit matrix from the schedule matrix. 
Given a schedule represented by the matrix in Figure~\ref{fig:cp}(b),
The result of the per-row suffix-or should be as Figure~\ref{fig:cp}(c), and the live-bit matrix should be as Figure~\ref{fig:cp}(d).

With the live-bit matrix, we can model the register pressure $z_k$ at every time-step \textit{k} 
\[z_k = \sum_{i=1}^{n} l_{ik}\times s_i\]
in which $s_i$ is a constant representing the output size of node \textit{i}.
In real implementation, we do not really introduce $z_k$ in the model. We only have
one integer variable Z which represents the MaxRP. For every time step \textit{k}, we add
the following inequality:
\[Z \geq \sum_{i=1}^{n} l_{ik}\times s_i\]

Finally, \textbf{the objective function is to minimize Z}.

Comparing to the integer-programming model defined in ~\cite{Govindarajan:2003:MRI:642768.642771}, this model is much simpler, only needs
3 $N\times N$ boolean variables and one integer object variable. We use Google OR-Tools to implement this model.
The per-basic-block DAGs are dumped from a GPU compiler that compiles shader programs used for rendering. 
We have tried both the integer-programming solver and the CP-SAT solver in the OR-Tools on those DAGs. The CP-SAT solver
turns out to be much more effective: solving much larger problems in the same amount of time. We have accumulated 4437 DAGs 
that the CP-SAT solver can prove it has found the optimal answers. The smallest DAGs have 48 nodes because we have filtered out DAGs with less than 48 nodes. 
We think the smaller problems are less interesting for this study. The largest DAG has 511 nodes, and the average is around 90 nodes.
In this process, we have encountered many large DAGs in which the CP-SAT solver only finds non-optimal solutions or cannot find a solution within the two-hour limit.
Those are excluded as well.

In the next section, we describe the heuristic algorithm for the min-reg scheduling. Using the optimal results obtained from the CP-SAT solver as the baseline,
We are able to understand how well those heuristics can achieve.

\section{Heuristic Scheduler}
\label{sec:scheduler}

Even with all recent advances in OR-Tools, the CP-SAT solver for min-reg scheduling is still too slow to use in a programming-language compiler.
From our experience, the instruction scheduling algorithm in production compilers should strive for a time-complexity of O(nlog(n)).
A time complexity of O($n^2$) starts having scaling issue with program size. A time complexity of O($n^3$) should be avoided whenever it is possible.
Graphics shader compilation also has its unique application environment: our compiler is used to compile thousands of shaders during game loading time. 
Slow game-loading time would be a severe user-experience issue, hence a serious competitive disadvantage for a GPU vendor. 
Also CP-SAT solver is very CPU-intensive. On a 4-core 8-thread I7 CPU, it constantly reaches over 90\% CPU utilization. 
3D Gaming itself tends to be both CPU and GPU intensive, so it would be hard to apply such CPU-hungry techniques on background compilation during game play. 

Our shader compiler has a typical code generation flow with separated instruction scheduling and register allocation phases. Instruction scheduling happens before
register allocation in order to exploit ILP. Instruction scheduler itself is also splitted into two phases: first min-reg scheduling then register-limited latency-hiding scheduling. 
At the beginning of the latency-hiding scheduling, we divide the basic block into sub-blocks based upon a worst-case estimation of register pressure 
that can potentially be created by greedy latency-hiding scheduling. This block-subdivision process is based upon the instruction sequence after the min-reg scheduling result: 
better min-reg scheduling result enables larger sub-blocks for latency-hiding scheduling. In the extreme case, if the register pressure is still high (above a given threshold) 
after min-reg scheduling, we simply skip the latency-hiding scheduling because it only exacerbates the spilling. For details on how to estimate the register pressure
and sub-divide the basic block before latency-hiding scheduling, please refer to the Section IV-B of ~\cite{IGC-CGO}. In this paper, we focus on how to
achieve good min-reg scheduling results with a polynomial-time algorithm.

\subsection{Sethi-Ullman on DAG}
\label{subsec:SU-DAG}

First,  we have tried to apply the generalized Sethi-Ullman algorithm directly on DAG, and see how well it can perform.
Algorithm~\ref{alg:sulist} shows a bottom-up list scheduler that uses ($MaxRP_i$ - $DefSZ_i$) to prioritize the ready queue.
In this paper, we call  ($MaxRP_i$ - $DefSZ_i$) the \textsf{SU-number} for convenience.
We call the base algorithm \textsf{bottom-up list-scheduling with the SU heuristic}. The time-complexity of updating the ready queue as a priority heap is O(log(n)) per step.
Therefore the overall complexity of this algorithm is O(nlog(n)).
Table~\ref{tab:SU0} shows the results on our test set. We use 3 metrics to compare the list-scheduling results against the optimal results:
\begin{itemize}
\item {The number of cases that reach the optimal.}
\item {The number of outlier cases in which the MaxRP found by the list-scheduler (LS) is 50\% over the optimal result.}
\item {The average ratio between the list-scheduling MaxRP and the optimal result.}
\end{itemize}

\begin{algorithm}
   \SetAlgoLined
   \KwData{\textbf{DAG}, a DAG to schedule}
   \KwData{N, the number of nodes in DAG}
   \KwData{$\textsf{MaxRP}_i$, the MaxRP of node i}
   \KwData{$\textsf{DefSZ}_i$, the output size of node i}
   \KwData{\textbf{Q}, the ready queue with the minimum($\textsf{MaxRP}_i$ - $\textsf{DefSZ}_i$) on top}
   \KwData{t, the time-step tracker}
         Compute $\textsf{MaxRP}_i$ for every node i in $\textbf{DAG}$\;
         Add all nodes without successors into \textbf{Q}\;
         t = N\;
         \While{\textbf{Q} not empty}{
             i = pop \textbf{Q}\;
             \textbf{SCHEDULE}[t] = i\;
             \For{every predecessor j of i}{
                 add j to \textbf{Q} if all successors of j are scheduled\;
             }
             t = t - 1\;
         }
\caption{bottom-up list scheduling with the SU heuristic}
\label{alg:sulist}
\end{algorithm}

\begin{table}[h]
   \begin{tabular}{|m{26mm}|m{15mm}|m{15mm}|m{15mm}|}
   \hline
   & Num Cases & Ave. Nodes Per Case & Ave Edge:Node Ratio \\ \hline
   Total Number of DAGs                                       & 4437  & 90.8 & 1.63 \\ \hline
   Cases that MaxRP by LS == Optimal MaxRP                   & 623  & 73.7 & 1.59  \\ \hline
   Cases that MaxRP by LS \textgreater{}= 1.5 * Optimal MaxRP & 1246  & 99.5 & 1.63 \\ \hline
   Average ratio of  MaxRP by LS over Optimal MaxRP           & 1.437 &  & \\ \hline
   \end{tabular}
\setlength{\abovecaptionskip}{12pt plus 3pt minus 2pt}
\caption{Generic Sethi-Ullman on DAG}
\label{tab:SU0}
\end{table}

In Table~\ref{tab:SU0}, we also collect two simple statistic features of DDGs: Column 3 is the average sizes of the DDGs; Column 4 is the average edge/node ratio fo the DDGs. The edge count includes all kinds of dependences that restricts instruction execution order. It shows that the heuristic approach tends to work better on the DDGs that are smaller and sparser. 
When we investigate the cases in which the SU heuristic does not work well, one observation is that the SU-numbers i.e. ($MaxRP_i$ - $DefSZ_i$), become less meaningful on DAGs with many nodes having multiple uses. For example, the ready queue tends to have multiple nodes with the equal SU-number.
So we should rely less on prioritizing the ready nodes with the SU-number. Without the reliable SU-numbers, which encode some sort of global register-usage information, 
We try to rely more on local information. For example, when a node becomes ready, we can immediately see that, for some node, adding it to the schedule does not
increase register pressure because either it has a large DefSize or its source operands are already in the live operand set, i.e. not the last uses of those operands.
If so, we schedule this kind of nodes immediately instead of adding them to the ready queue. In essence, we make the algorithm more like a greedy algorithm. 
We call this the \textsf{RP-reduction} heuristic. Table~\ref{tab:SU2} shows the result of adding the \textsf{RP-reduction} heuristic. 
We get some significant improvements: 107 more optimal cases and 122 less outlier cases when comparing to the generic Sethi-Ullman result.  The average ratio is reduced by 1.9 percentage point.  
\begin{table}[h]
   \begin{tabular}{|l|l|}
   \hline
   Total Number of DAGs                                       & 4437  \\ \hline
   Cases that MaxRP by LS == Optimal MaxRP                   & 730   \\ \hline
   Cases that MaxRP by LS \textgreater{}= 1.5 * Optimal MaxRP & 1124   \\ \hline
   Average ratio of  MaxRP by LS over Optimal MaxRP           & 1.418 \\ \hline
   \end{tabular}
\setlength{\abovecaptionskip}{12pt plus 3pt minus 2pt}
\caption{Sethi-Ullman and RP-reduction heuristic}
\label{tab:SU2}
\end{table}

For better clarity, we present the improved list-scheduler in Algorithm~\ref{alg:sulist2}, which incorporates the RP-reduction  heuristics. 
Because it is a bottom-up scheduling algorithm, an operand's live-range starts when we schedule the last use of that operand ("last" in the sense of execution order).
An operand's  live-range ends when we schedule its defining node. We use $LiveTS_i$ to denote the time-step at which the last use of node \textit{i} is scheduled. 
We can use $LiveTS$ array to evaluate the register-pressure change caused by a node-scheduling action.
Notice that, in our algorithm,  if a node that becomes ready and does not increase register-pressure, we schedule it immediately. 
By doing so,  we avoid the extra complexity of searching such instructions from the ready queue. The time-complexity of the algorithm should still be at O(nlog(n))
\begin{algorithm}
   \SetAlgoLined
   \KwData{\textbf{DAG}, a DAG to schedule}
   \KwData{N, the number of nodes in DAG}
   \KwData{$\textsf{MaxRP}_i$, the MaxRP of node i}
   \KwData{$\textsf{DefSZ}_i$, the output size of node i}
   \KwData{$\textsf{LiveTS}_i$, the time-step for the last use of node i}
   \KwData{\textbf{Q},  the ready queue with the minimum($\textsf{MaxRP}_i$ - $\textsf{DefSZ}_i$) on top}
  \KwData{\textbf{W}, the set of nodes selected to be scheduled}
   \KwData{t, the time-step tracker}
         Compute $\textsf{MaxRP}_i$ for every node i in $\textbf{DAG}$\;
         Initialize $\textsf{LiveTS}_i$ to -1 for every node in $\textbf{DAG}$\;
         Add all nodes without successors into \textbf{Q}\;
         t = N\;
         \While{\textbf{Q} not empty}{
             i = pop \textbf{Q}\;
             add i to \textbf{W}\;
             \While{\textbf{W} not empty}{
                 i = pop \textbf{W}\;
                 \textbf{SCHEDULE}[t] = i\;
                 \For{every source-operand j of i}{
                     $\textsf{LiveTS}_j$ = max(t, $\textsf{LiveTS}_j$)\;
                 }
                 \For{every predecessor j of i}{
                     \If{all successors of j are scheduled}{
                        d = $\textsf{DefSZ}_i$ if $\textsf{LiveTS}_i$ >0 else 0\;
                        \For{every source-operand m of j}{
                            d = (d - $\textsf{DefSZ}_m$) if $\textsf{LiveTS}_m$ < 0\;
                        }
                        \eIf{$d\geq 0$}{
                            add j to \textbf{W}\;
                        }{
                            add j to \textbf{Q}\;
                        }
                     }
                 }
                 t = t - 1\;
             }
         }
\caption{Improved SU scheduler with RP-reduction heuristics}
\label{alg:sulist2}
\end{algorithm}

Looking at the cases that our algorithm is still not effective, we notice that issues seem always related to those nodes with multiple uses. So we also try to adjust the $MaxRP_i$: 
dividing it by the number of its uses. Table~\ref{tab:SU3} shows the result after adding the $MaxRP_i$ adjustment. Looks like this adjustment is quite effective in improving Sethi-Ullman scheduling on DAG. 
111 more optimal cases and 49 less outlier cases. More important, the average ratio is reduced by 4.2 percentage point. 
Therefore we think that the treatment of those multi-use nodes is critical for further improvement. That is what we are going to address in the next sub-section.

\begin{table}[h]
   \begin{tabular}{|l|l|}
   \hline
   Total Number of DAGs                                       & 4437  \\ \hline
   Cases that MaxRP by LS == Optimal MaxRP                   & 841   \\ \hline
   Cases that MaxRP by LS \textgreater{}= 1.5 * Optimal MaxRP & 1075   \\ \hline
   Average ratio of  MaxRP by LS over Optimal MaxRP           & 1.376 \\ \hline
   \end{tabular}
\setlength{\abovecaptionskip}{12pt plus 3pt minus 2pt}
\caption{With both RP-reduction and $MaxRP_i$ adjustment}
\label{tab:SU3}
\end{table}

\subsection{Scheduling DAG in clusters}
\label{subsec:SU-DAG-clustering}

\begin{figure}[h]
\center
\includegraphics[width=0.3\textwidth]{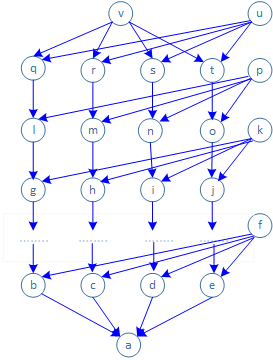}
\caption{A tightly-coupled DAG}
\label{fig:cluster}
\end{figure}

Graphics shaders tend to have the following kind of sequence:
\begin{lstlisting}
u = ALUOp(...)
v = ALUOp(...)
(r, g, b) = TextureSample(u, v)
x = ALUOp(r, ...)
y = ALUOp(g, ...)
z = ALUOp(b, ...)
\end{lstlisting}
GPU instruction-set requires our register allocator must allocate (r, g, b) together into consecutive registers, (u, v) together in consecutive registers. 
However, ALU operations on those values have to be scalarized. When we build a data-dependence graph for this example, 
that texture-sample instruction has 3 uses. Live range (u, v) also has 2 uses: first as a partial-write input of "v= ...", then as the input of texture-sample. 
The GPU programs we deal with tend to have a lot of multi-use nodes due to prevailing usage of short-vector operands. 
Therefore, the clustering situation shown in Figure~\ref{fig:cluster} is fairly typical. So we use  Figure~\ref{fig:cluster} as the motivating example 
for the next improvement we describe in this subsection. In this example, there are 4 chains of computation: (q, l, g, ..., b), (r, m, h, ..., c), (s, n, i, ..., d), and (t, o, j, ..., e).
However, these 4 chains of computation are tied together by those multi-use nodes (v, u, p, k, ..., f).
If we apply the generic SU-scheduling. Algorithm~\ref{alg:sulist}, to this example, since it favors the node with the minimum SU number at each step.
It schedules these four chains separately, i.e. first schedule (b, ..., g, l, q) then (c, ..., h, m, r) and so on.
Scheduling in this order, we end up with very high MaxRP because the live ranges of all the multi-use nodes (f, ..., k, p, u, v) are overlapped.
The right strategy is to schedule (b, c, d, e) together then schedule f immediately, schedule (g, h, i, j) together then schedule k, and so on.
Scheduling in this order, we only need 5 registers at each level. We call a set of nodes that share some common source operands as a cluster.
Therefore, in Figure~\ref{fig:cluster}, (b, c, d, e) is a cluster, (g, h, i, j) is a cluster and so on.

We have modified our scheduler to recognize those nodes that form a cluster and try to schedule them together.
Algorithm~\ref{alg:cluster} is built upon Algorithm~\ref{alg:sulist2}. So we have skipped many repeated lines in the algorithm description, and focus on the cluster-handling part.
The cluster-forming starts with the top node on the ready queue. We find other unscheduled nodes in DAG that have common source operands with the top node.
These nodes form a cluster. If all the nodes in the cluster are in the ready queue, we schedule the cluster together.
If any node in the cluster is not in the ready queue, we follow this node's successors, and its successor's successor and so on(since we are doing bottom-up scheduling) 
until we find a node in the ready queue.
We schedule that node next. Essentially, we hold the unscheduled cluster, and try to schedule those nodes that prevent the cluster from becoming ready. 
Adding the clustering heuristic definitely increases the time-complexity from O(nlog(n)) to O($n^2$) due to the extra work for finding the ready node that unblocks the cluster.
\begin{algorithm}
   \SetAlgoLined
   \KwData{all items declared in \textbf{Alg2} }
   \KwData{\textsf{cluster}, the set of nodes sharing common sources}
         Initialization parts are the same as \textbf{Alg2}\;
         \While{\textbf{Q} not empty}{
             i = the top of \textbf{Q}\;
             add i to \textsf{cluster}\;
             add other unscheduled nodes that share common source operands with the cluster\;
             \eIf{all nodes in cluster are ready}{
                 add nodes in cluster to \textbf{W}\;
                 remove nodes in cluster from \textbf{Q}\;
             }{
                j =  a node in cluster that is not ready\;
                k = find a node in the ready queue that blocks j\;
                add k to \textbf{W}\;
                remove k from \textbf{Q}\;
             }
             \While{\textbf{W} not empty}{
                  Loop body inside While-W is the same as \textbf{Alg2}\;
             }
         }
\caption{Scheduling with Clustering Heuristic}
\label{alg:cluster}
\end{algorithm}
Looking at the results after adding the clustering heuristic in Table~\ref{tab:SU4}, we can see that scheduling by clusters is very important for achieving good min-reg result 
on those GPU programs. Compared with Table~\ref{tab:SU3}, all three indicators are improved significantly.
The average ratio decreases from 1.376 to 1.171 (20 percentage points). The number of outlier cases decreases from 1075 to 266. The number of optimal cases increases from 841 to 1427.
Looking at the statistics on two graph features, the cluster heuristic can handle larger and denser graphs. The outlier cases still tend to be larger, however, there is no longer a correlation between the edge/node ratio and the resulting MaxRP. 
\begin{table}[h]
   \begin{tabular}{|m{26mm}|m{15mm}|m{15mm}|m{15mm}|}
   \hline
   & Num Cases & Ave. Nodes Per Case & Ave Edge:Node Ratio \\ \hline
   Total Number of DAGs                                       & 4437  & 90.8 & 1.63 \\ \hline
   Cases that MaxRP by LS-cluster == Optimal MaxRP                   & 1427 & 78.1 & 1.64   \\ \hline
   Cases that MaxRP by LS-cluster \textgreater{}= 1.5 * Optimal MaxRP & 266 & 115.9 & 1.58   \\ \hline
   Average ratio of  MaxRP by LS-cluster over Optimal MaxRP           & 1.171  &  & \\ \hline
   \end{tabular}
\setlength{\abovecaptionskip}{12pt plus 3pt minus 2pt}
\caption{Clustering Heuristic plus RP-reduction}
\label{tab:SU4}
\end{table}

Finally, in Table~\ref{tab:SU5}, we try to add the $MaxRP_i$ adjustment to the cluster-scheduling algorithm as well. Looks like the clustering heuristic and the $MaxRP_i$ adjustment are not quite compatible. Combining them ends up with worse results than Table~\ref{tab:SU4}.

\begin{table}[]
   \begin{tabular}{|l|l|}
   \hline
   Total Number of DAGs                                               & 4437  \\ \hline
   Cases that MaxRP by LS-cluster == Optimal MaxRP                   & 1136   \\ \hline
   Cases that MaxRP by LS-cluster \textgreater{}= 1.5 * Optimal MaxRP & 273    \\ \hline
   Average ratio of  MaxRP by LS-cluster over Optimal MaxRP                   & 1.194 \\ \hline
   \end{tabular}
\setlength{\abovecaptionskip}{12pt plus 3pt minus 2pt}
\caption{Clustering plus MaxRP adjustment and RP-reduction}
\label{tab:SU5}
\end{table}

\section{Applying Min-Reg Scheduling In Intel Graphics Compiler}
\label{sec:exp}

We have implemented those heuristic min-reg scheduling algorithms described in the previous section in Intel Graphics Compiler(IGC)~\cite{IGC-CGO}.
Our target GPU is Intel ARC\texttrademark  GPU, which has 128 registers per thread. 
IGC uses a two-phase scheduling approach before the register allocation: first min-reg scheduling then register-limited latency scheduling.
The purpose of min-reg scheduling is to set a good base for register-limited latency scheduling. Therefore the performance contribution of min-reg scheduling is indirect.
Also IGC employs an empirical way to filter out the outlier effect of min-reg scheduling: it compares the min-reg schedule with the original sequence. 
If the register pressure is higher after min-reg scheduling than the original sequence, it simply uses the original sequence as the input for latency scheduling.
The original sequence is just another point in the solution space that we can utilize even though we do not know its quality.
However incorporating this empirical technique does make it even more subtle to demonstrate the contribution of min-reg scheduling to the final performance.

In this section, we start with one case we have encountered in a real application to demonstrate how more effective min-reg scheduling can result in better final code sequence.

\begin{figure}[h]
\center
\includegraphics[width=0.5\textwidth]{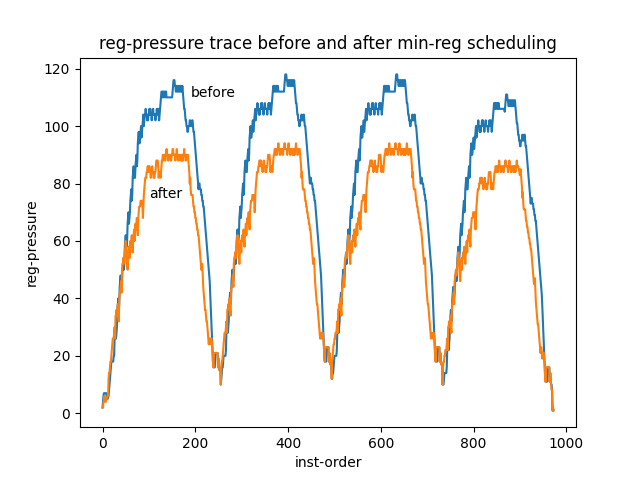}
\caption{Register pressure before and after min-reg scheduling}
\label{fig:minreg-plot}
\end{figure}

\begin{figure}[h]
\center
\includegraphics[width=0.5\textwidth]{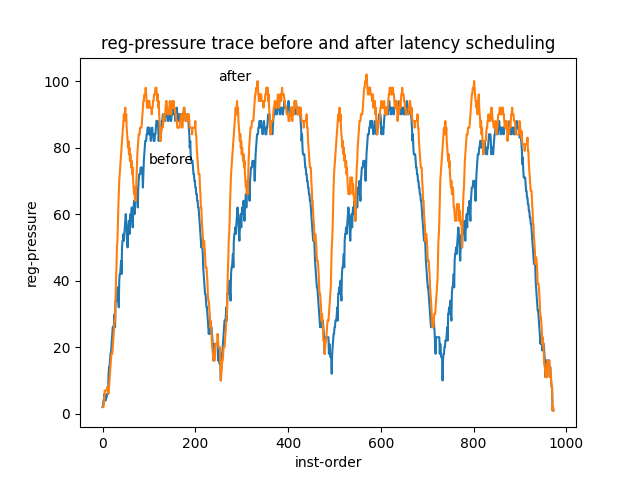}
\caption{Register pressure before and after latency scheduling }
\label{fig:latency-plot}
\end{figure}

The example we selected is a compute-shader that has only one large basic block (after a loop is fully unrolled 4 times). 
Figure \ref{fig:minreg-plot} plots the register-pressure change along the instruction sequence before and after the min-reg scheduling. 
As it is shown, the clustering-based min-reg scheduling is able to reduce the $MaxRP$ in this case from 118 registers to 94 registers. 
That gives some room to allow latency-hiding scheduling. Figure \ref{fig:latency-plot} plots the register-pressure change along 
the instruction sequence before and after the latency-hiding scheduling. The $MaxRP$ after latency-hiding scheduling 
is 102. The sharp peaks in the plot after latency-hiding scheduling are the results of scheduling multiple long-latency memory operations 
at the beginning of every loop iteration. After register allocation and post-RA instruction scheduling, IGC estimates the time to
execute the final sequence is 2350 cycles. For the same case, if we apply the generic Sethi-Ullman scheduling, we are not able to 
reduce the $MaxRP$ of the input sequence. The $MaxRP$ of 118 is above the threshold that we allow for applying latency-hiding scheduling.
Therefore the latency-hiding scheduling step is skipped. Without pre-RA latency-hiding scheduling, after register allocation and post-RA scheduling,
IGC estimates the execution time for that final sequence is 4359 cycles. Note that IGC's cycle estimation is after considering 
the latency-hiding effect of simultaneous multithreading on GPU. In some other more extreme cases,  we have observed that, 
without effective min-reg scheduling, register allocator needs to insert spill code, which incurs even larger cycle penalties. 
On the other hand,  effective min-reg scheduling successfully avoids spilling, and leads to a final sequence with much short cycle estimation.

As we have indicated with the example, we evaluate two min-reg scheduling algorithms, which are described as Algorithm~\ref{alg:minreg1} and Algorithm~\ref{alg:minregcluster}.
Also Note that, in IGC, we only apply these min-reg scheduling algorithms under the following conditions:
\begin{itemize}
\item The original $MaxRP$ for a block is greater than 64.
\item The $MaxRP$ is not at the block-entry or block-exit because, otherwise, instruction reordering within the basic block does not help.
\end{itemize}
  
The baseline we use for evaluation is the IGC without any min-reg scheduling before latency-hiding scheduling.

\begin{algorithm}
\SetAlgoLined
\KwData{\textbf{B}, a basic block to schedule}
      \textsf{OldMaxRP} = Compute register pressure \textsf{rp} of \textbf{B}\;
      \textsf{NewMaxRP} = \textbf{GenericSethiUllmanScheduling}(\textbf{B})\;
      \If{$\textsf{NewMaxRP} < \textsf{OldMaxRP}$}{
          Commit min-reg schedule\;
     }
\caption{Generic Sehti-Ullman Scheduling in IGC}
\label{alg:minreg1}
\end{algorithm}

\begin{algorithm}
\SetAlgoLined
\KwData{\textbf{B}, a basic block to schedule}
      \textsf{OldMaxRP} = Compute register pressure \textsf{rp} of \textbf{B}\;
      \textsf{NewMaxRP} = \textbf{SethiUllmanSchedulingWithClusters}(\textbf{B})\;
      \eIf{$\textsf{NewMaxRP} < \textsf{OldMaxRP}$}{
           Commit min-reg schedule\;
      }{
            \textsf{NewMaxRP} = \textbf{GenericSethiUllmanScheduling}(\textbf{B})\;
            \If{$\textsf{NewMaxRP} < \textsf{OldMaxRP}$}{
                  Commit min-reg schedule\;
            }
      }
\caption{Sethi-Ullman Scheduling With Clustering in IGC}
\label{alg:minregcluster}
\end{algorithm}

We conduct our experiment on a set of compute-shaders and pixel-shaders that we have collected from popular games running on Windows.
This set of shaders are newer than those shaders that we have found optimal solution using the CP-SAT solver. There are a total of 3383 shaders.
When we break shaders into basic blocks, there are a total of 53049 basic blocks. 

\begin{figure}[h]
\center
\includegraphics[width=0.5\textwidth]{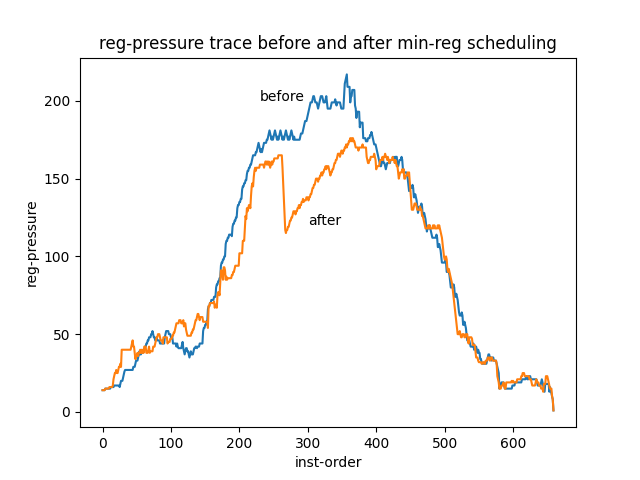}
\caption{Another example before and after min-reg scheduling}
\label{fig:minreg-morespills}
\end{figure}

First let's look at some summary data for big picture:
\begin{itemize}
\item Applying the baseline compiler(without any min-reg scheduling), 873 out of 3383 shaders have spill code inserted by register allocator.
\item Applying the compiler with the Algorithm~\ref{alg:minreg1}, 2561 out of 53049 basic blocks have picked the SU scheduling output, then 823 shaders end up with spill code.
\item Applying the compiler with the Algorithm~\ref{alg:minregcluster}, 2854 out of 53049 blocks have picked the clustering-based scheduling output. Beyond that, 344 blocks have picked the SU scheduling output. Finally 815 shaders end up with spill code.
\end{itemize}
These results indicate that, even though the vast majority of blocks do not need min-reg scheduling, there are cases in which min-reg scheduling is essential because it can reduce spills.
These results also demonstrate that our clustering heuristic is an improvement to the generic Sethi-Ullman algorithm because it can reduce register pressure on more blocks (2854 versus 2516), and it reduces the spilling shaders from 823 to 815. The summary data only counts the shaders with spills versus the shaders without spills. It does not show the cases in which spills are reduced (but not to zero) or the cases in which there is no spill but latency hiding is improved because of adding min-reg scheduling. 
We can see these effects by examining the cycle estimation from IGC for each shader. IGC reports an estimated cycle for each shader by doing a weighted sum of the estimated cycles for all basic blocks in the shader(weighted by the loop-nesting level).
\begin{figure}[h]
\center
\includegraphics[width=0.5\textwidth]{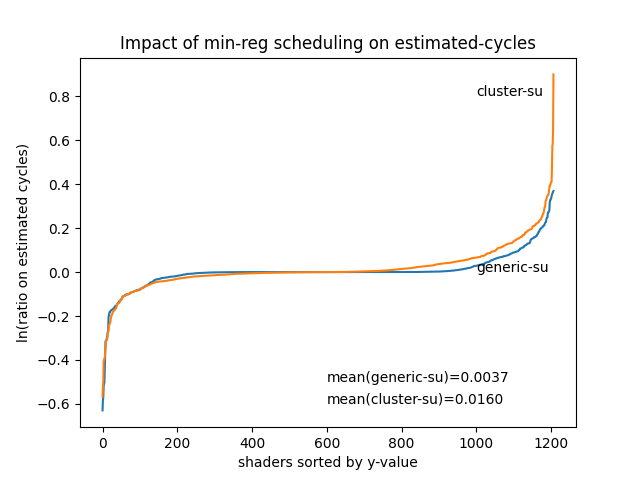}
\caption{Impact on estimated cycles by Algorithm~\ref{alg:minreg1} and ~\ref{alg:minregcluster}}
\label{fig:s-curve}
\end{figure}
Figure \ref{fig:s-curve} plots two s-curves. Each s-curve demonstrates the effect of one min-reg scheduling algorithm in terms of cycle estimation when comparing to the baseline result, which is the cycle estimation without any min-reg scheduling. The Y-axis value is formulated as \textit{ln(baseline-cycles/min-reg-cycles)}. The X-axis points represent shader cases that are sorted by their corresponding Y-value. Since the cases are sorted by the Y-values, the points further away from the origin are the cases with improvement, and the points closer to the origin are the cases with degradation.  There are only about 1200 shaders because we have filtered out those cases that are not affected by any sort of min-reg scheduling. We can draw several conclusions from these two curves:
\begin{itemize}
\item The net effect of min-reg scheduling on cycle reduction is positive because the means of both s-curves are positive.
\item The Algorithm~\ref{alg:minregcluster} is more effective than the Algorithm~\ref{alg:minreg1} because its mean is more positive.
\item There are cases in which we have reduced register pressure in some blocks, yet ended up with higher cycle estimation.
\end{itemize}
We have examined one case in which min-reg scheduling leads to higher cycle estimation. Figure \ref{fig:minreg-morespills} shows the register-pressure traces before and after min-reg scheduling for that case. As it is shown, the clustering-based min-reg scheduling is able to reduce the $MaxRP$ from 217 to 176.  somehow, our register allocator ends up generating more spill code for the trace after min-reg scheduling. We attribute this result to the less-than-perfect heuristics used by our graph-coloring algorithm. Another potential issue we have realized is that, when we estimate register pressure during instruction scheduling, we have omitted many architectural restrictions in register assignment. These restrictions can cause unused fragments in the register file, however they are not modeled during instruction scheduling. In other words, we believe that there are places that need improvement in both our instruction scheduler and register allocator, which may fix many of those negative cases that get exposed by this experiment.

\section{Related Work}
\label{sec:related}

In this section, we discuss prior works in two related areas: First, works on using integer programming or constraint programming to model instruction scheduling and register allocation problems. Second, works on polynomial-time heuristic scheduling algorithms for reducing register pressures.

On the first topic, Govindarajan \textit{et al}~\cite{Govindarajan:2003:MRI:642768.642771} proposed an integer-programming model to find the min-reg sequence. We have described the difference between their work and ours in the previous sections. Wilken \textit{et al}~\cite{Wilken2000IntegerProgramming} proposed an integer-programming model for instruction scheduling. Malik \textit{et al}~\cite{Malik2006ConstraintProgramming} proposed a constraint-programming model for instruction scheduling.
Both of these models are targeted to minimize the overall execution time of a basic block. Similar to our model, they all model the ordering constraints based upon data dependences. Different from our model, they model latency and function-unit constraints but do not model register constraints. Lozano \textit{et al}~\cite{Lozano2019unison} proposed a constraint programming model that solves the instruction scheduling and register allocation problem together. Their model should have the most comprehensive formulation of the constraints. Also they have integrated their solver into the LLVM compiler to generate code for several CPUs. Comparing to their work, we are studying a simpler
graph optimization problem. The problem is simpler in terms of the number of constraints we need to model, however these are all NP-complete problems. Also we do not use our CP-SAT solver in actual code generation due to compilation-time requirements in our application. We only use our constraint programming model to  help develop and evaluate polynomial-time approximation algorithms.

On the second topic of developing polynomial-time approximation algorithm, there are also works that try to solve the problem by integrating instruction scheduling with the register allocation ~\cite{Goodman:1988:CSR:55364.55407,Palem:1993:STI:155183.155190,Pinter:1993:RAI:155090.155114,Hsu1989OnTM}. 
The heuristic min-reg algorithm proposed by Govindarajan \textit{et al}~\cite{Govindarajan:2003:MRI:642768.642771} can be classified into this category as well because it involves building an interference graph among node-lineages in a DAG then applying the graph-coloring algorithm. Due to the complexity of implementing integrated approaches, we have not seen much adoption in the modern production compilers including LLVM.
For scheduling large basic blocks, Goodman and Hsu proposed the heuristic of picking instructions that reduces register pressure in ~\cite{Goodman:1988:CSR:55364.55407}. In this paper, we have adopted their idea in our RP-reduction heuristic. 
Even though some heuristics that are similar to ours may have already been tried in production or research compilers, 
We have not seen studies published to evaluate their effectiveness for the min-reg scheduling problem. That is one question that this paper tries to answer. In this paper, we demonstrate that a combination of several heuristics can achieve a more effective min-reg schedule than the generic Sethi-Ullman algorithm. Furthermore, by comparing our heuristic algorithms with the CP-SAT solver, we provide better insights on how much headroom there is for further improvement.

\section{Conclusion}
\label{sec:conclusion}
The min-reg scheduling problem may have other applications beyond the CPU and GPU code generation.
For example, in the machine-learning compiler, there may be similar problems for scheduling data-flow graphs to minimize the intermediate buffers.
The constraint-programming model we have proposed may become more applicable in those applications where compilation-time becomes less of an issue.
Regarding the next step of this work, it is natural to consider exploring the machine-learning techniques for more effective min-reg scheduler,
especially, after we can obtain a large set of optimal answers using the SP-SAT solver.

\section{ACKNOWLEDGMENT}
\label{sec:related}
We thank Wei-Chung Hsu and anonymous reviewers for reviewing this paper and giving many constructive suggestions.
We also thank Wei Pan who designed and implemented the original instruction scheduler in IGC,
and all the colleagues who worked on IGC over the past years.

\bibliographystyle{ieeetran}
\bibliography{cp-sat-min-reg}

\begin{thebibliography}{10}
\providecommand{\url}[1]{#1}
\csname url@samestyle\endcsname
\providecommand{\newblock}{\relax}
\providecommand{\bibinfo}[2]{#2}
\providecommand{\BIBentrySTDinterwordspacing}{\spaceskip=0pt\relax}
\providecommand{\BIBentryALTinterwordstretchfactor}{4}
\providecommand{\BIBentryALTinterwordspacing}{\spaceskip=\fontdimen2\font plus
\BIBentryALTinterwordstretchfactor\fontdimen3\font minus
  \fontdimen4\font\relax}
\providecommand{\BIBforeignlanguage}[2]{{%
\expandafter\ifx\csname l@#1\endcsname\relax
\typeout{** WARNING: IEEEtran.bst: No hyphenation pattern has been}%
\typeout{** loaded for the language `#1'. Using the pattern for}%
\typeout{** the default language instead.}%
\else
\language=\csname l@#1\endcsname
\fi
#2}}
\providecommand{\BIBdecl}{\relax}
\BIBdecl

\bibitem{IGC-CGO}
A.~Chandrasekhar, G.~Chen, P.-Y. Chen, W.-Y. Chen, J.~Gu, P.~Guo, S.~H.~P.
  Kumar, G.-Y. Lueh, P.~Mistry, W.~Pan, T.~Raoux, and K.~Trifunovic, ``Igc: The
  open source intel graphics compiler,'' in \emph{2019 IEEE/ACM International
  Symposium on Code Generation and Optimization (CGO)}, 2019, pp. 254--265.

\bibitem{Sethi:1970:GOC:321607.321620}
\BIBentryALTinterwordspacing
R.~Sethi and J.~D. Ullman, ``The generation of optimal code for arithmetic
  expressions,'' \emph{J. ACM}, vol.~17, no.~4, pp. 715--728, Oct. 1970.
  [Online]. Available: \url{http://doi.acm.org/10.1145/321607.321620}
\BIBentrySTDinterwordspacing

\bibitem{appel1987generalizations}
A.~W. Appel and K.~J. Supowit, ``Generalizations of the sethi-ullman algorithm
  for register allocation,'' \emph{Software: Practice and Experience}, vol.~17,
  no.~6, pp. 417--421, 1987.

\bibitem{ortools}
\BIBentryALTinterwordspacing
L.~Perron and V.~Furnon, ``Or-tools,'' Google. [Online]. Available:
  \url{https://developers.google.com/optimization/}
\BIBentrySTDinterwordspacing

\bibitem{Aho:1977:CGE:321992.322001}
\BIBentryALTinterwordspacing
A.~V. Aho, S.~C. Johnson, and J.~D. Ullman, ``Code generation for expressions
  with common subexpressions,'' \emph{J. ACM}, vol.~24, no.~1, pp. 146--160,
  Jan. 1977. [Online]. Available:
  \url{http://doi.acm.org/10.1145/321992.322001}
\BIBentrySTDinterwordspacing

\bibitem{Govindarajan:2003:MRI:642768.642771}
\BIBentryALTinterwordspacing
R.~Govindarajan, H.~Yang, J.~N. Amaral, C.~Zhang, and G.~R. Gao, ``Minimum
  register instruction sequencing to reduce register spills in out-of-order
  issue superscalar architectures,'' \emph{IEEE Trans. Comput.}, vol.~52,
  no.~1, pp. 4--20, Jan. 2003. [Online]. Available:
  \url{https://doi.org/10.1109/TC.2003.1159750}
\BIBentrySTDinterwordspacing

\bibitem{Wilken2000IntegerProgramming}
\BIBentryALTinterwordspacing
K.~Wilken, J.~Liu, and M.~Heffernan, ``Optimal instruction scheduling using
  integer programming,'' in \emph{Proceedings of the ACM SIGPLAN 2000
  Conference on Programming Language Design and Implementation}, ser. PLDI
  '00.\hskip 1em plus 0.5em minus 0.4em\relax New York, NY, USA: Association
  for Computing Machinery, 2000, p. 121–133. [Online]. Available:
  \url{https://doi.org/10.1145/349299.349318}
\BIBentrySTDinterwordspacing

\bibitem{Malik2006ConstraintProgramming}
A.~M. Malik, J.~McInnes, and P.~van Beek, ``Optimal basic block instruction
  scheduling for multiple-issue processors using constraing programming,'' in
  \emph{2006 18th IEEE International Conference on Tools with Artificial
  Intelligence (ICTAI'06)}, 2006, pp. 279--287.

\bibitem{Lozano2019unison}
\BIBentryALTinterwordspacing
R.~C.~n. Lozano, M.~Carlsson, G.~H. Blindell, and C.~Schulte, ``Combinatorial
  register allocation and instruction scheduling,'' \emph{ACM Trans. Program.
  Lang. Syst.}, vol.~41, no.~3, jul 2019. [Online]. Available:
  \url{https://doi.org/10.1145/3332373}
\BIBentrySTDinterwordspacing

\bibitem{Goodman:1988:CSR:55364.55407}
\BIBentryALTinterwordspacing
J.~R. Goodman and W.-C. Hsu, ``Code scheduling and register allocation in large
  basic blocks,'' in \emph{Proceedings of the 2Nd International Conference on
  Supercomputing}, ser. ICS '88.\hskip 1em plus 0.5em minus 0.4em\relax New
  York, NY, USA: ACM, 1988, pp. 442--452. [Online]. Available:
  \url{http://doi.acm.org/10.1145/55364.55407}
\BIBentrySTDinterwordspacing

\bibitem{Palem:1993:STI:155183.155190}
\BIBentryALTinterwordspacing
K.~V. Palem and B.~B. Simons, ``Scheduling time-critical instructions on risc
  machines,'' \emph{ACM Trans. Program. Lang. Syst.}, vol.~15, no.~4, pp.
  632--658, Sep. 1993. [Online]. Available:
  \url{http://doi.acm.org/10.1145/155183.155190}
\BIBentrySTDinterwordspacing

\bibitem{Pinter:1993:RAI:155090.155114}
\BIBentryALTinterwordspacing
S.~S. Pinter, ``Register allocation with instruction scheduling,'' in
  \emph{Proceedings of the ACM SIGPLAN 1993 Conference on Programming Language
  Design and Implementation}, ser. PLDI '93.\hskip 1em plus 0.5em minus
  0.4em\relax New York, NY, USA: ACM, 1993, pp. 248--257. [Online]. Available:
  \url{http://doi.acm.org/10.1145/155090.155114}
\BIBentrySTDinterwordspacing

\bibitem{Hsu1989OnTM}
W.-C. Hsu, C.~N. Fischer, and J.~R. Goodman, ``On the minimization of
  loads/stores in local register allocation,'' \emph{IEEE Transactions on
  Software Engineering}, vol.~15, pp. 1252--1260, 1989.

\end{thebibliography}



\end{document}